\documentclass[]{aastex631}
\usepackage{amsmath}
\usepackage{graphicx}
\usepackage{booktabs}

\usepackage{hyperref}
\makeatletter
\def\@drafttext{}
\def\@journalinfo{}
\makeatother

\usepackage{float}
\usepackage{stfloats}
\shorttitle{Torlakcik Catalog: SETI Target Avoidance}
\shortauthors{Torlakcik}
\usepackage{flafter}
\usepackage{placeins}

\begin{document}

\setkeys{Gin}{width=0.35\textwidth}
\title{Where Not to Look: A Parametric Avoidance Model for SETI Target Selection}

\author{Sahin Torlakcik}
\correspondingauthor{Sahin Torlakcik}
\email{torlakciksahin@gmail.com}
\affiliation{Ankara Atat\"urk High School, Ankara, T\"urkiye}

\begin{abstract}
We present a simple, rule-based filter for SETI target selection that flags stars unlikely to host complex life and produces an audit-ready exclusion catalog. Using seven stellar parameters, including age, metallicity, and multiplicity, the model excludes roughly half of a 1.74 million-star Gaia DR3 sample, retaining 777,835 high-priority targets, mainly G and K dwarfs. Age and metallicity dominate the rejections.
Importantly, using Gaia's age upper bounds instead of point estimates saves
355,086 stars from exclusion. A comparison of empirical and synthetic
proxies shows that while the overall exclusion rate is robust, individual
target assignments change significantly; for instance, the commonly used
RUWE indicator flags 2.7$\times$ more binaries than Gaia's own non-single-star
flag. Cross-matching with the Breakthrough Listen target list reveals a 56.5\%
exclusion rate, highlighting the complementary nature of habitability-driven
and proximity-driven surveys. The catalog, pipeline, and a generalized
community tool are publicly available.
\end{abstract}
% abstract iyi oldu

\section{Introduction}
\label{sec:intro}

The Search for Extraterrestrial Intelligence (SETI) has to allocate finite observing
time across an enormous stellar search space. Current surveys such as Breakthrough
Listen \citep{Isaacson2017} target tens of thousands of stars, while the full Gaia
catalog contains over one billion sources. Even restricting attention to stars with
complete astrophysical parameter coverage yields millions of candidates, far
exceeding the capacity of any single observational program. So efficient target
prioritization is arguably just as important as the observations themselves.

Earlier works like HabCat \citep{Turnbull2003} established systematic de-emphasis criteria for SETI target selection, identifying stellar properties that reduce the likelihood of hosting communicating civilizations. These criteria include minimum stellar age, photometric stability, appropriate spectral class, metallicity, and dynamical stability of the habitable zone. Subsequent large-scale radio surveys, most notably the Breakthrough Listen (BL) program \citep{Isaacson2017, Gajjar2021, Czech2021}, prioritize nearby, bright stars to maximize sensitivity, as do wide-field Galactic plane surveys \citep{MarcyTellis2024}, but apply relatively permissive astrophysical filtering. A generalized, catalog-agnostic filtering framework based explicitly on habitability criteria, one that can be applied to any stellar dataset and provides clear, traceable reasons for each decision, does not yet exist in the literature.

We present such a framework. The \textit{Torlakcik Catalog} formalizes established
SETI de-emphasis criteria into a transparent, rule-based parametric avoidance
function that accepts stellar parameters as input and returns binary accept/reject
decisions accompanied by machine-readable reason codes. The model is intentionally
rule-based rather than probabilistic, prioritizing interpretability and computational
lightness for resource-constrained workflows. We apply the model to a Gaia DR3 sample
($N = 1{,}742{,}306$), report exclusion statistics, perform sensitivity analysis on
all threshold values, compare empirical versus synthetic proxy performance, and
cross-match results with existing SETI survey target lists.

The contributions of this work are: (a) a formal, auditable rule-based filter with
reason codes suitable for scheduling pipelines; (b) an uncertainty-aware age criterion
using Gaia DR3 age upper bounds; (c) an empirical test on 1.74 million Gaia DR3 stars;
(d) a quantitative comparison between empirical and synthetic proxy approaches that
shows the insensitivity of fractional flux error as a variability proxy; (e)
sensitivity analysis characterizing how exclusion rates respond to threshold variation;
(f) cross-matching with the Breakthrough Listen target list; and (g) open-source release
of both the paper-specific analysis pipeline, the generalized community software,
and the full exclusion catalog.

Although distance remains a primary driver of observational feasibility in SETI surveys
\citep{Isaacson2017, Gajjar2021, Czech2021}, the present work deliberately focuses on
astrophysical habitability criteria that are independent of distance. For practical context,
the median distance of the $777{,}835$ retained high-priority targets in our sample with
reliable parallax ($\varpi/\sigma_\varpi > 5$) is $382$~pc. This distance range lies well
within the sensitivity volume of current and planned radio SETI facilities such as the Green
Bank Telescope, Parkes/MeerKAT, and the Very Large Array, confirming that the
Torlakcik Catalog can serve as a pre-filter for existing target lists without requiring
additional distance-based cuts.
%to do adding marcy tellis
\section{Data}
\label{sec:data}

\subsection{Gaia DR3 Query}
We queried Gaia Data Release 3 \citep{GaiaCollaboration2023} via the ESA TAP
service. The query joins \texttt{gaiadr3.astrophysical\_parameters},
\texttt{gaiadr3.astrophysical\_parameters\_supp}, and
\texttt{gaiadr3.gaia\_source}, with a left join on
\texttt{gaiadr3.vari\_summary} for photometric variability flags. We applied
non-null filters on \texttt{teff\_gspphot}, \texttt{mh\_gspphot},
\texttt{mass\_flame\_spec}, and \texttt{age\_flame\_spec}, and required reliable
parallax ($\varpi/\sigma_\varpi > 5$) to support distance calculations.
The query also retrieves \texttt{age\_flame\_spec\_upper} for the
uncertainty-aware age criterion described in Section~\ref{sec:model},
\texttt{ruwe} for the synthetic multiplicity proxy, and
\texttt{phot\_g\_mean\_flux} with \texttt{phot\_g\_mean\_flux\_error} for the
synthetic variability proxy, as described in Section~\ref{sec:proxy}.
Additionally, \texttt{phot\_g\_mean\_mag} and \texttt{bp\_rp} are retrieved
for constructing the color-magnitude diagram presented in
Section~\ref{sec:hrd}.
The initial query returned $1{,}754{,}135$ stars satisfying the non-null
parameter filters; applying the parallax reliability criterion removed
$11{,}829$ sources ($0.67\%$), yielding a final sample of $N = 1{,}742{,}306$
stars. The negligible attrition rate indicates that the parallax filter
introduces minimal selection bias in this query. The complete TAP query is
reproduced in Appendix~\ref{app:query}.

\subsection{Spectral Type Classification}

Spectral types were inferred from \texttt{teff\_gspphot} using standard effective
temperature boundaries \citep{Gray2009}: $T_\mathrm{eff} \geq 30{,}000$~K
$\rightarrow$ O; $10{,}000$--$29{,}999$~K $\rightarrow$ B; $7{,}500$--$9{,}999$~K
$\rightarrow$ A; $6{,}600$--$7{,}499$~K $\rightarrow$ F0--F4;
$6{,}000$--$6{,}599$~K $\rightarrow$ F5--F9; $5{,}200$--$5{,}999$~K $\rightarrow$
G; $3{,}700$--$5{,}199$~K $\rightarrow$ K; $< 3{,}700$~K $\rightarrow$ M.

\subsection{Distance Estimation}
\label{sec:distance}

Stellar distances are estimated as $d = 1000/\varpi$ (pc), where $\varpi$ is the
Gaia DR3 parallax in milliarcseconds. This simple inversion is standard practice
for samples with high parallax signal-to-noise; for our $\varpi/\sigma_\varpi > 5$
criterion, the Lutz--Kelker bias \citep{LutzKelker1973, Smith2003} is
$< 2\%$ for the median star in the sample and remains below $5\%$ for
$\varpi/\sigma_\varpi > 3$. We therefore adopt the naive inversion throughout
this work, noting that the parallax reliability cut itself constitutes a
non-probabilistic distance selection.

We acknowledge that a more rigorous treatment would employ the Bayesian
likelihood framework of \citet{BailerJones2021}, which infers distances from
parallaxes using an exponentially decreasing space density prior. This approach
accounts for the asymmetric and non-Gaussian distance uncertainties that arise
from parallax inversion, and was adopted by \citet{Czech2021} for the MeerKAT
SETI target sample. However, because our avoidance criteria operate on
intrinsic stellar parameters (mass, age, metallicity, spectral type) rather
than on distance directly, and because the median fractional parallax
uncertainty in our sample is $\sim 10\%$ (corresponding to $\sim 20\%$ distance
uncertainty at most), the choice of distance estimator does not affect the
exclusion decisions. Distance enters the analysis only in the summary statistics
(median distance of retained stars) and the FOV coverage calculations
(Table~\ref{tab:fov}), where the $1/\varpi$ approximation introduces a
systematic bias well within the precision required for those calculations.

\subsection{Relationship to the Gaia Catalog of Nearby Stars}
\label{sec:gcns}

The Gaia Catalog of Nearby Stars \citep[GCNS;][]{GCNS2021} provides a
volume-complete sample of 331,312 stars within 100~pc, selected from Gaia EDR3
with rigorous parallax quality cuts and a probabilistic distance framework.
We cross-matched the GCNS against our sample using \texttt{source\_id},
recovering 29,258 GCNS members with complete astrophysical parameter coverage.
Of these, 16,822 ($57.5\%$) are retained by our model, a higher fraction
than the full catalog ($44.6\%$), reflecting the relative age maturity and
higher median metallicity of the nearby stellar population. The primary
exclusion drivers for nearby stars are low metallicity (R4: 7,704) and young
age (R2: 3,884), with the remaining criteria contributing smaller fractions
(R1: 613; R3: 588; R5: 2,308; R6: 1,654; R7: 3). The GCNS
selection function (parallax-based with \citealt{BailerJones2021} distances)
and our own ($\varpi/\sigma_\varpi > 5$ with $1/\varpi$ distances) adopt
different but complementary approaches to distance determination; the
substantial overlap in recovered stars confirms the consistency of both
approaches for the nearby sample.

\section{Model}
\label{sec:model}

\subsection{Avoidance Criteria}

The parametric avoidance function evaluates each star against seven criteria.
A star is excluded if it satisfies \textit{any} of the following:

\vspace{0.5em}
\noindent\textbf{R1: Mass:} $M > 1.5\,M_\odot$. Stars above this mass have
main sequence lifetimes shorter than $\sim$2~Gyr \citep{KippenhahnWeigert1990},
severely limiting the time available for complex life to develop. Consistent
with \citet{Turnbull2003}.

\vspace{0.3em}
\noindent\textbf{R2: Age:} $\tau_\mathrm{upper} < 3$~Gyr. Complex life on Earth
required $\sim$3--4~Gyr to emerge \citep{Knoll2021}; systems younger than this
threshold have not had enough time for comparable evolutionary processes
\citep{Turnbull2003}. Gaia DR3 stellar ages carry substantial uncertainties
\citep{GaiaCollaboration2023}; we therefore apply this threshold to the
\texttt{age\_flame\_spec\_upper} column rather than the point estimate, ensuring
that only stars whose \textit{upper} age bound falls below 3~Gyr are excluded.
This retains 355,086 stars that a hard cut on point estimates would discard.
Figure~\ref{fig:age_comparison} illustrates the impact of this choice.

\vspace{0.3em}
\noindent\textbf{R3: Spectral type:} O, B, A, or F0--F4. O and B stars have
lifetimes $< 1$~Gyr \citep{HansenKawaler1994}, while A and early-F stars produce
UV radiation environments hostile to surface life \citep{Scalo2007}.

\vspace{0.3em}
\noindent\textbf{R4: Metallicity:} $[\mathrm{Fe/H}] < -0.4$. Giant planet
formation efficiency drops sharply below this threshold \citep{Fischer2005,
Johnson2010}, reducing the expected frequency of habitable terrestrial planets.

\noindent\textbf{R5: Multiplicity:} \texttt{non\_single\_star} $\geq 1$.
We exclude all stars flagged as non-single in Gaia DR3. The
\texttt{non\_single\_star} field indicates the presence of at least one
Non-Single Star solution (astrometric, spectroscopic, or eclipsing) in the
Gaia NSS tables. Because this flag does not reliably distinguish wide binaries
from dynamically more complex triple or higher-order multiples, we
conservatively exclude all flagged sources on the grounds that gravitational
perturbations in multiple systems can destabilize planetary orbits
\citep{HolmanWiegert1999, Quarles2020}. We note that the exclusion of
binary and multiple systems is also standard practice in SETI survey
design: \citet{Czech2021} applied an astrometric $\chi^2$ cut to their
Gaia DR2 sample specifically to remove binary systems, noting that
``binary stars have poor distance estimates.''

\vspace{0.3em}
\noindent\textbf{R6: Photometric variability:}
\texttt{range\_mag\_g\_fov} $> 0.01$~mag, or

\texttt{phot\_variable\_flag} = \texttt{VARIABLE}. Variable stellar irradiance
affects planetary climate stability and may erode atmospheric biosignatures
\citep{Kasting1993}.

\vspace{0.3em}
\noindent\textbf{R7: M-dwarf activity:} M-type stars with
\texttt{in\_vari\_rotation\_modulation} or \texttt{in\_vari\_short\_timescale}
flagged True. Chromospherically active M dwarfs produce frequent high-energy
flares that can erode planetary atmospheres \citep{France2016, Howard2018}.
Chromospherically quiet M dwarfs are retained as candidates, consistent with the
growing consensus that such stars merit SETI attention \citep{Shields2016}.

\subsection{Decision Function and Reason Codes}

\begin{figure}[!ht]
    \centering
    \includegraphics[width=\columnwidth, height=0.85\textheight, keepaspectratio]{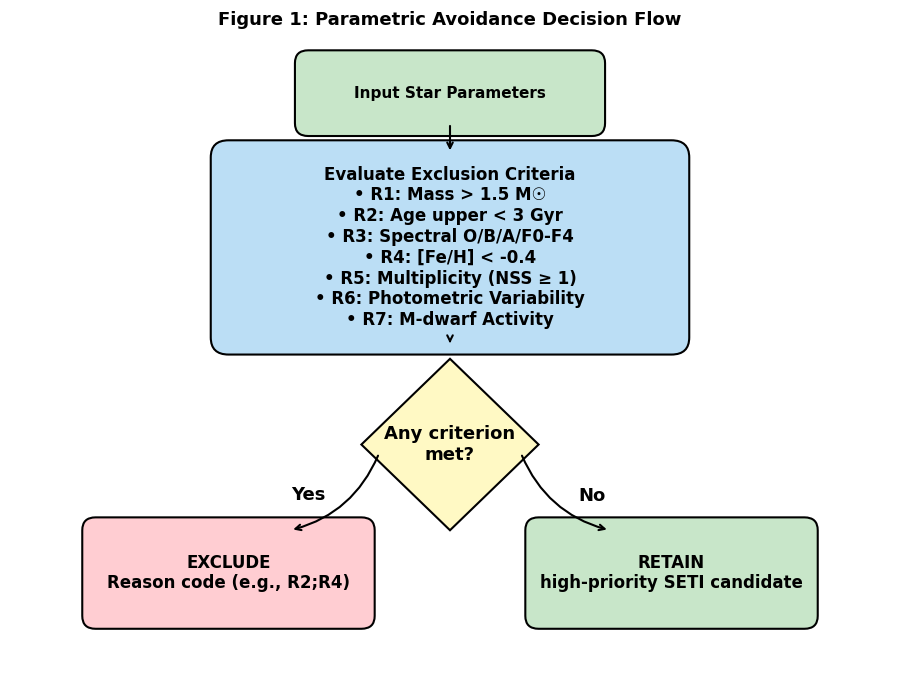}
    \caption{Decision flow of the parametric avoidance function. Stars failing
    any threshold check receive the corresponding reason code and are flagged
    as excluded. Stars passing all checks are retained as SETI candidates.}
    \label{fig:decision_flow}
\end{figure}

Each star is evaluated against all seven criteria simultaneously. Stars
failing any criterion are assigned an EXCLUDE flag and a semicolon-delimited
reason code string (e.g., \texttt{R2;R4} for a young, metal-poor star). Stars
passing all checks are flagged RETAIN. The decision flow is illustrated in
Figure~\ref{fig:decision_flow}. The function is catalog-agnostic: any stellar
dataset containing the required parameter columns can be used as input.

\section{Results}
\label{sec:results}

\subsection{Overall Exclusion Statistics}

Applied to $N = 1{,}742{,}306$ Gaia DR3 stars with reliable parallax, the model
excludes 964,471 stars (55.4\%) and retains 777,835 (44.6\%) as high-priority SETI candidates. Per-criterion counts are given in Table~\ref{tab:results}.

\begin{table}[!h]
\centering
\caption{Per-criterion exclusion counts ($N = 1{,}742{,}306$). Stars may
satisfy multiple criteria; counts are not mutually exclusive.}
\label{tab:results}
\begin{tabular}{llrr}
\toprule
Code & Criterion & $N_\mathrm{excl}$ & Fraction \\
\midrule
R1 & Mass $> 1.5\,M_\odot$       & 295,019 & 16.9\% \\
R2 & Age upper $< 3$~Gyr         & 502,305 & 28.8\% \\
R3 & Spectral O/B/A/F0--F4       & 130,082 &  7.5\% \\
R4 & $[\mathrm{Fe/H}] < -0.4$    & 504,318 & 29.0\% \\
R5 & Multiplicity (NSS $\geq 1$) & 120,213 &  6.9\% \\
R6 & Phot.\ variability          &  68,796 &  3.9\% \\
R7 & Active M dwarf              &      84 & $<0.1$\% \\
\midrule
   & \textbf{Total excluded} & \textbf{964,471}  & \textbf{55.4\%} \\
   & \textbf{Total retained} & \textbf{777,835}  & \textbf{44.6\%} \\
\bottomrule
\end{tabular}

\end{table}

\subsection{Per-Criterion Breakdown}

\begin{figure}[!ht]
    \centering
    \includegraphics[width=\columnwidth]{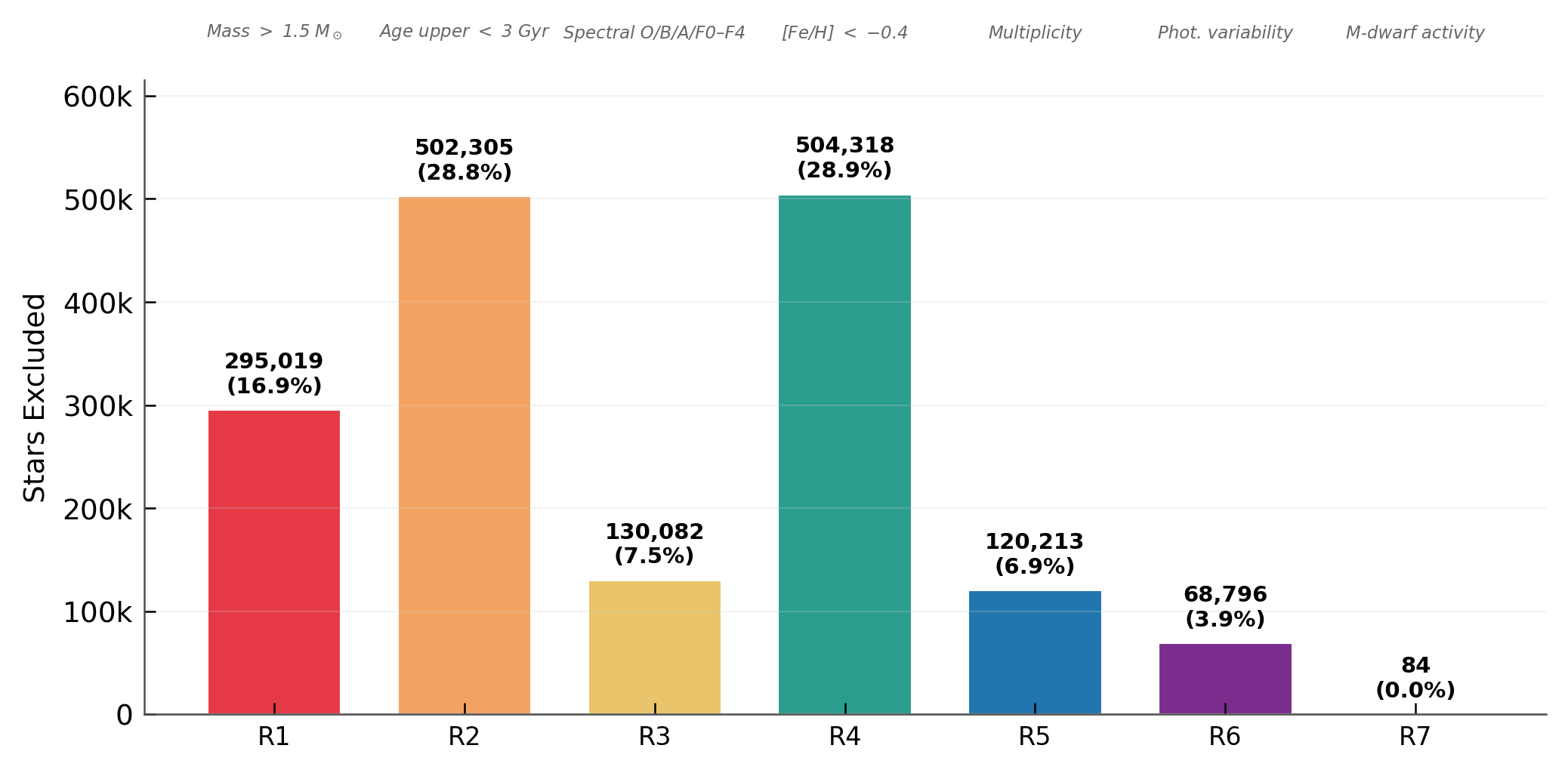}
    \caption{Number of stars excluded by each avoidance criterion
    ($N = 1{,}742{,}306$). The age threshold (R2) and metallicity threshold
    (R4) are co-dominant exclusion drivers. The age count is substantially reduced
    relative to a point-estimate cut because we apply the criterion to the Gaia DR3
    age upper bound.}
    \label{fig:criteria}
\end{figure}

Figure~\ref{fig:criteria} shows the number of stars excluded by each criterion.
Age (R2) and metallicity (R4) are co-dominant drivers, each excluding $\sim$29\%
of the sample. M-dwarf activity (R7) flags only 84 stars, which reflects the conservative
nature of the Gaia variability classification flags used for this criterion.

\subsection{Age Criterion: Uncertainty-Aware Filtering}

\begin{figure}[!h]
    \centering
    \includegraphics[width=0.8\columnwidth]{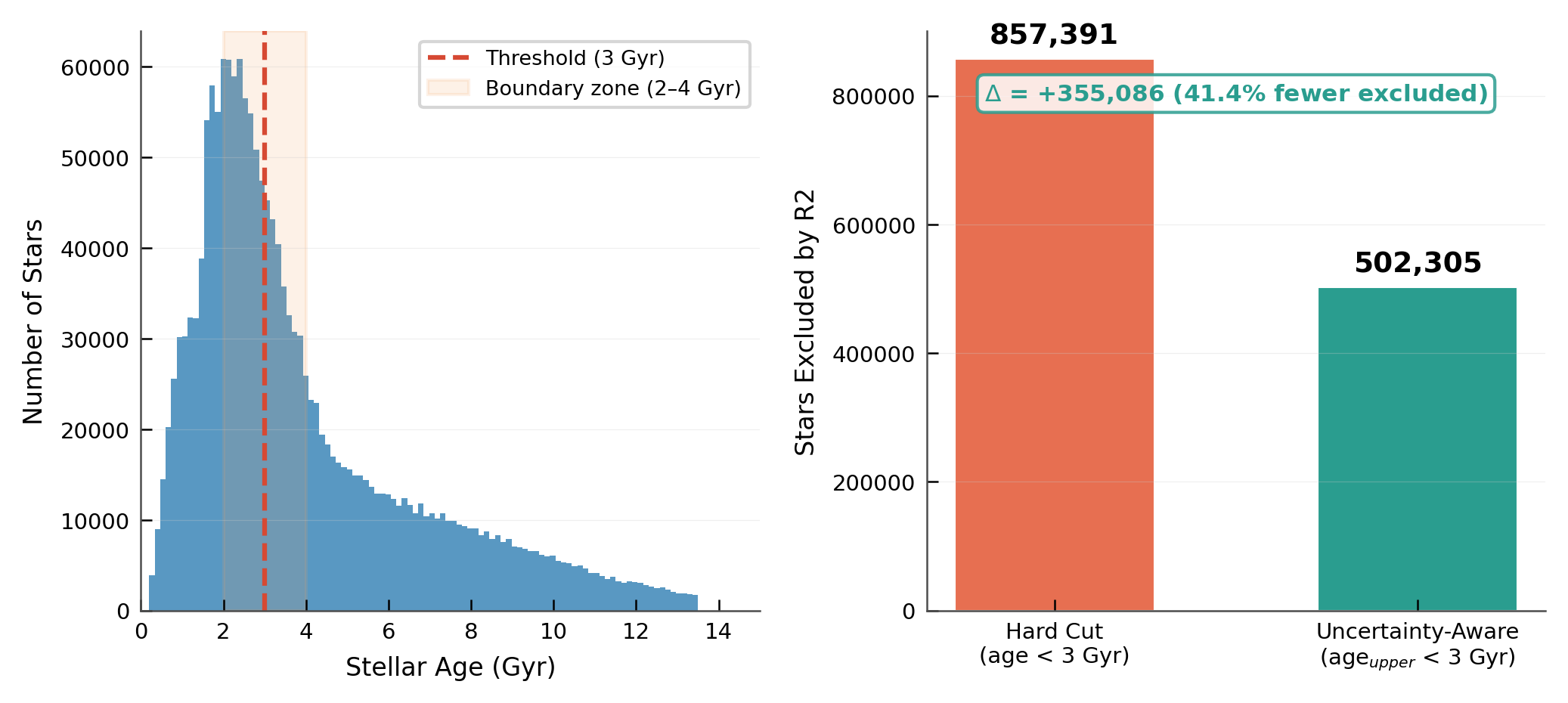}
    \caption{Comparison of age-driven exclusions under the original hard-cut
    criterion (\texttt{age\_flame\_spec} $< 3$~Gyr; red) and the uncertainty-aware
    criterion (\texttt{age\_flame\_spec\_upper} $< 3$~Gyr; green).
    The uncertainty-aware approach retains 355,086 additional stars.}
    \label{fig:age_comparison}
\end{figure}

Figure~\ref{fig:age_comparison} compares the number of stars excluded under the
original hard cut ($\tau < 3$~Gyr, $N_\mathrm{excl} = 857{,}391$) and the revised
uncertainty-aware criterion ($\tau_\mathrm{upper} < 3$~Gyr, $N_\mathrm{excl} = 502{,}305$).
The difference of 355,086 stars represents a $41\%$ reduction in age-driven
exclusions. These stars have point-estimate ages below 3~Gyr but age upper bounds
that are consistent with the threshold, placing them in a transition zone
($39.8\%$ of the sample lies in the 2--4~Gyr boundary zone). Given the large
systematic uncertainties in Gaia photometric ages \citep{GaiaCollaboration2023}, the conservative choice of applying the threshold to the upper bound is physically motivated. We note that the apparent excess of young stars in the age distribution (more at 0--5~Gyr than 5--10~Gyr) likely reflects a systematic bias in the FLAME algorithm rather than a real star-formation rate enhancement (see Section~\ref{sec:limitations}).

\subsection{Spectral Type Breakdown}

\begin{figure}[!h]
    \centering
    \includegraphics[width=\columnwidth]{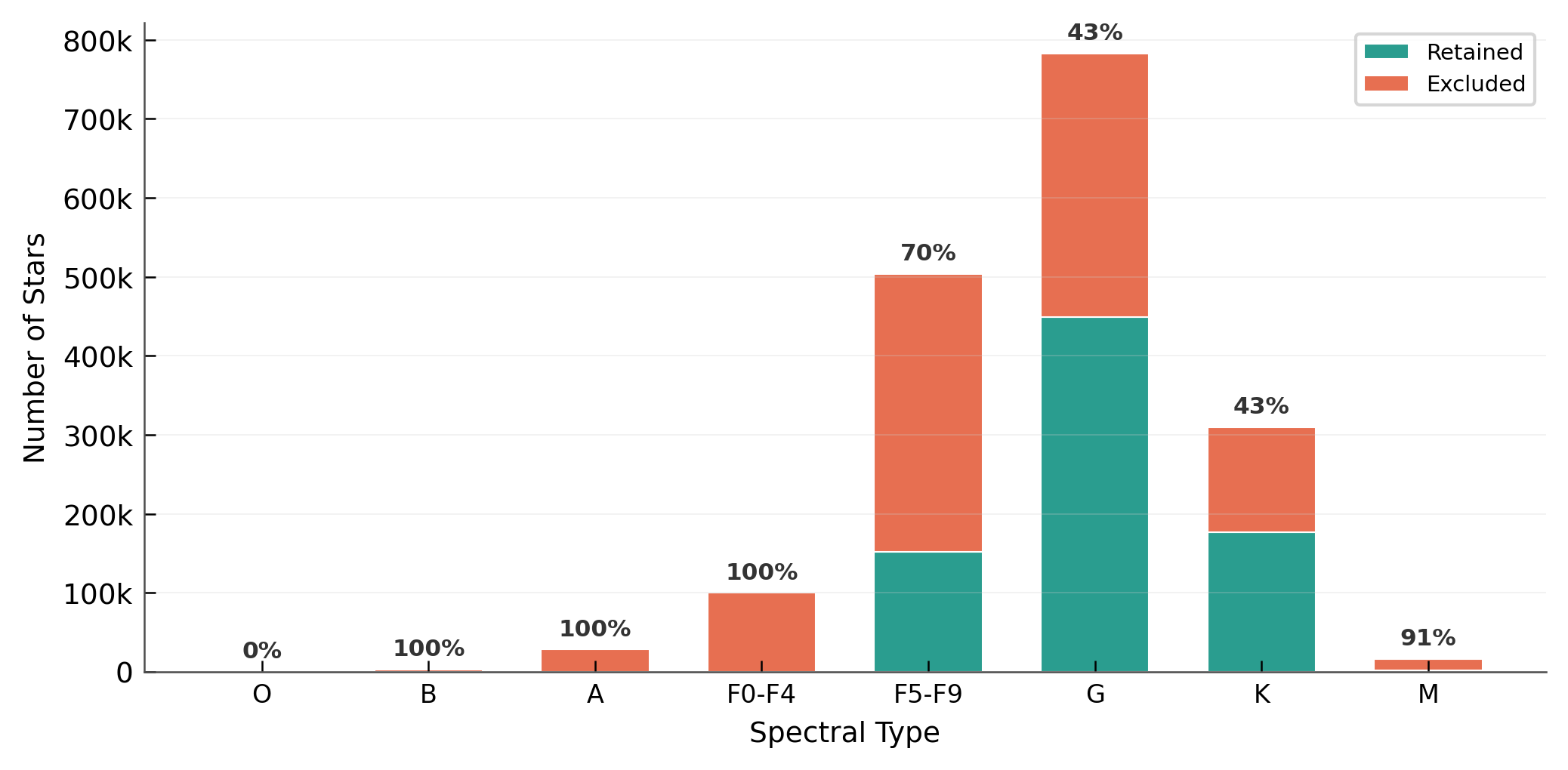}
    \caption{Retained (green) and excluded (red) star counts by spectral type.
    B, A, and F0--F4 stars are excluded at 100\% by the spectral criterion.
    G and K dwarfs dominate the retained population.}
    \label{fig:spectype}
\end{figure}

Figure~\ref{fig:spectype} shows retained and excluded counts by spectral type.
B, A, and F0--F4 stars are excluded at 100\% by criterion R3 (O stars are
absent in this parallax-filtered sample). F5--F9 stars are excluded at $70\%$,
driven mainly by R2 and R4. G and K stars are excluded at $43\%$, and
M dwarfs at $91\%$, both dominated by the age and metallicity criteria. The
lower G and K exclusion fractions compared to the original unfiltered sample
reflect the relative age maturity of nearer G/K stars with precise parallaxes.

\subsection{Hertzsprung-Russell (HR) Diagram}
\label{sec:hrd}

\begin{figure}[!h]
    \centering
    \includegraphics[width=\columnwidth]{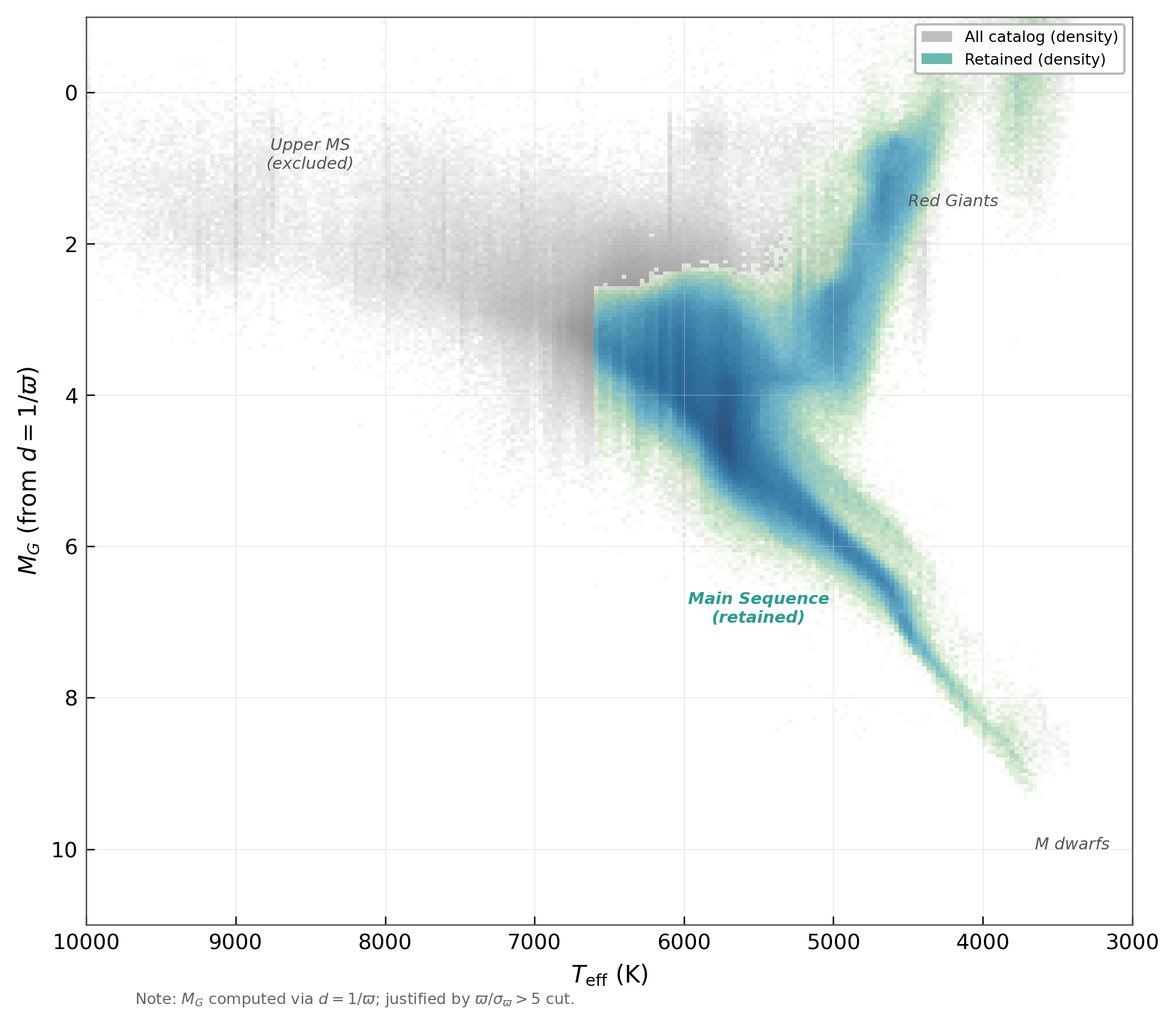}
    \caption{Hertzsprung-Russell (HR) diagram ($T_\mathrm{eff}$ vs. $M_G$) for the sample. Grey: excluded stars. Teal: retained candidates. The retained population populates the main sequence from late-F to M-type stars and includes a prominent red giant branch.}
    \label{fig:hrd}
\end{figure}

Figure~\ref{fig:hrd} presents the Hertzsprung-Russell (HR) diagram of the sample, with excluded stars shown as a light grey background and retained candidates overlaid in teal. The retained population cleanly populates the main sequence from late-F ($T_{\text{eff}} \approx 6500$ K) through K-type stars, with a sparse extension into the M-dwarf region. Additionally, a prominent branch of red giant stars is retained in the sample.
Two primary features of the diagram merit attention. First, the excluded population includes essentially all stars on the upper main sequence ($T_{\text{eff}} > 6500$ K, corresponding to O, B, and A types). Second, the retained population traces a clear main sequence from late-F through M-type stars, alongside the aforementioned red giant branch. This selection demonstrates the effective isolation of these specific lower-mass stellar populations while discarding the hotter upper main sequence.

\subsection{Empirical vs.\ Synthetic Proxy Comparison}
\label{sec:proxy}

\begin{figure}[!h]
    \centering
    \includegraphics[width=\columnwidth]{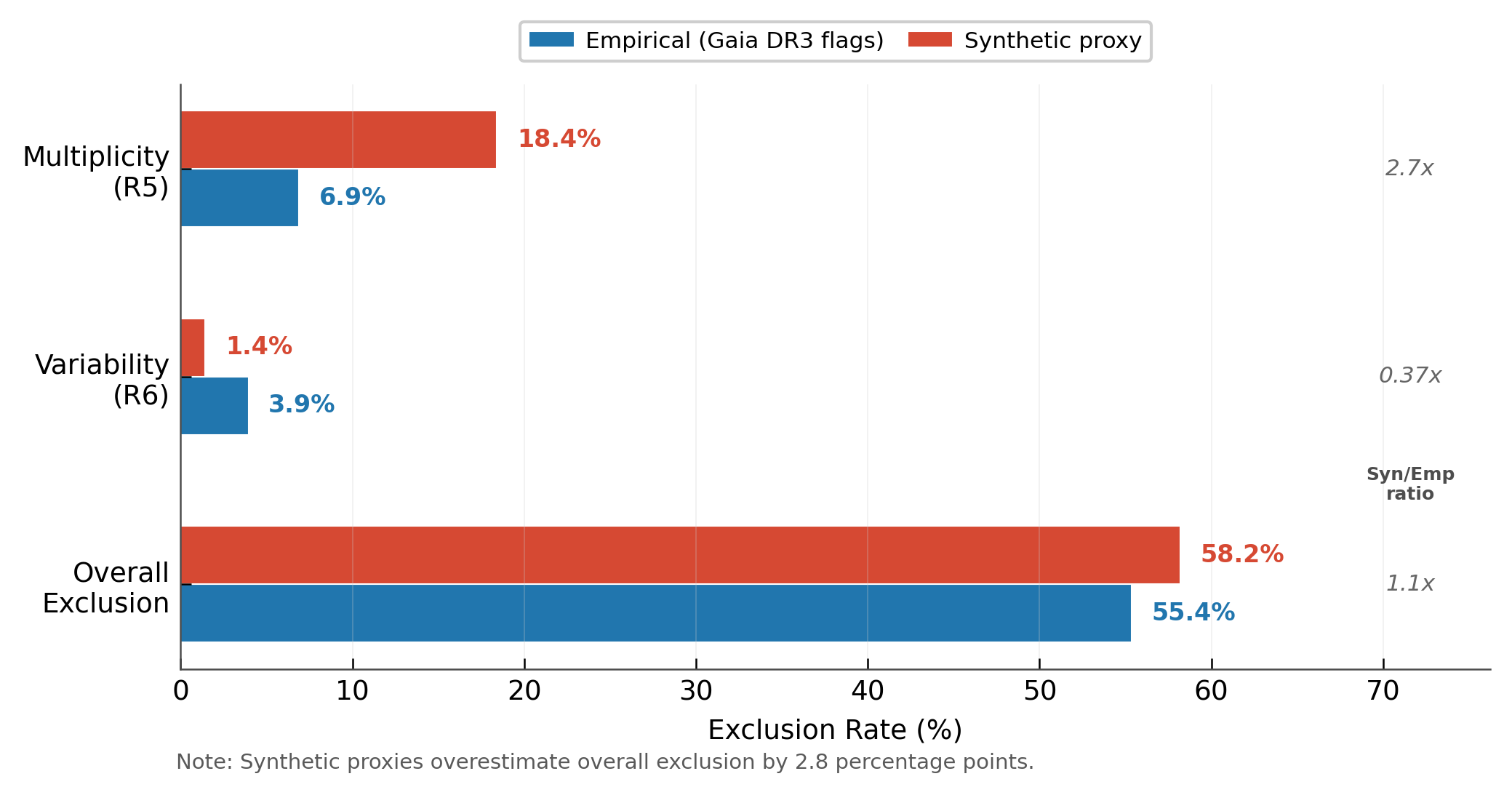}
    \caption{Exclusion rates under empirical Gaia DR3 flags (blue) and
    synthetic proxies (red) for multiplicity, variability, and overall
    exclusion. Synthetic proxies overestimate overall exclusion by 2.8~percentage points,
    with RUWE-driven R5 overestimation partially offset by
    $\sigma_G/G$-driven R6 underestimation.}
    \label{fig:synthetic}
\end{figure}

To assess how robust our exclusion criteria are to the choice of input
data products, we compare the empirical Gaia DR3 classification flags used
in our baseline model against synthetic proxies derived from independently
measured quantities. This comparison serves two purposes: it quantifies how
sensitive the exclusion results are to the specific flags available in a
given catalog, and it tests whether physically motivated proxies can
reproduce the empirical results in catalogs lacking dedicated classification
flags. Results are summarized in Table~\ref{tab:proxy} and
Figure~\ref{fig:synthetic}.

\begin{table}[!h]
\centering
\caption{Empirical vs.\ synthetic proxy comparison. The synthetic R5 proxy
(RUWE $> 1.4$) is more inclusive than the Gaia NSS flag, while the synthetic
R6 proxy ($\sigma_G/G > 0.001$) is substantially less sensitive than empirical
variability flags. Overall exclusion rates agree to within 2.8~percentage points.}
\label{tab:proxy}
\begin{tabular}{lrrr}
\toprule
Criterion & Empirical & Synthetic & Ratio \\
\midrule
R5: Multiplicity & 6.9\% & 18.4\% (RUWE$>$1.4) & 2.7$\times$ \\
R6: Variability  & 3.9\% & 1.4\% ($\sigma_G/G>$0.001) & 0.36$\times$ \\
Overall exclusion & 55.4\% & 58.2\% & --- \\
\bottomrule
\end{tabular}
\end{table}

\subsubsection{Multiplicity: RUWE vs.\ Non-Single-Star Flag}

For multiplicity (R5), the empirical proxy is the
\texttt{non\_single\_star} flag ($\geq 1$), indicating at least one
Non-Single Star solution in the Gaia NSS tables. The synthetic proxy is the
Renormalized Unit Weight Error (RUWE), where RUWE $> 1.4$ indicates
astrometric excess noise consistent with unresolved multiplicity
\citep{Lindegren2021}.

The RUWE $> 1.4$ threshold flags 2.7 times more stars than the
\texttt{non\_single\_star} flag ($18.4\%$ vs.\ $6.9\%$). This discrepancy
reflects the different selection functions: the NSS flag requires a
successfully fitted orbital solution, a stringent criterion that only
captures well-characterized binaries with sufficient orbital coverage and
signal-to-noise ratio. RUWE, by contrast, captures any astrometric
perturbation, including unresolved companions too faint or too close for
orbital fitting, as well as stars with significant spot-induced photocenter
motion \citep{Belokurov2020}. From a SETI perspective, the RUWE threshold
provides a more conservative exclusion: any star exhibiting astrometric
noise consistent with a companion is potentially subject to dynamical
complications in its planetary system, whether or not a full orbital
solution has been achieved.

\subsubsection{Variability: Fractional Flux Error vs.\ Gaia Classification}
For photometric variability (R6), the empirical proxy combines
\texttt{range\_mag\_g\_fov} $> 0.01$~mag with the
\texttt{phot\_variable\_flag} = \texttt{VARIABLE} classification. The
synthetic proxy is the fractional G-band flux error
$\sigma_G/G \equiv$ \texttt{phot\_g\_mean\_flux\_error} $/$
\texttt{phot\_g\_mean\_flux}, a measure of photometric precision that might
intuitively correlate with variability.

The fractional flux error turns out to be an insensitive proxy for photometric
variability. At a threshold of $\sigma_G/G > 0.001$ ($0.1\%$), only $1.4\%$
of stars are flagged, compared to $3.9\%$ under the empirical flags. Raising
the threshold to $\sigma_G/G > 0.01$ ($1.0\%$) captures merely $0.01\%$ of
the sample, and a $5\%$ threshold captures zero stars. This insensitivity is
a direct consequence of Gaia's extraordinary photometric precision: the median
$\sigma_G/G$ in our sample is $0.020\%$ (mean $0.025\%$), and only $1.45\%$
of stars exceed $0.1\%$. The fundamental issue is that $\sigma_G/G$ measures
the \textit{precision of the mean flux estimate}, not the \textit{amplitude
of flux variation}. A well-observed variable star with many Gaia transits can
have a precisely determined mean flux even if its instantaneous flux varies
substantially between observations. Conversely, \texttt{range\_mag\_g\_fov}
directly measures the peak-to-peak amplitude of G-band magnitude variation,
and \texttt{phot\_variable\_flag} incorporates Gaia's dedicated variability
classification pipeline \citep{Eyer2023}. These empirical products are
inherently more sensitive to stellar variability than the fractional error on
the mean.

This finding has implications beyond the present work. The fractional
photometric error, while an intuitive and easily computed quantity, should
not be used as a variability proxy in Gaia-based studies without explicit
validation against survey-specific classification flags.

\subsubsection{Overall Exclusion and Partial Cancellation}

Despite the large individual differences in R5 ($+11.5$~percentage points) and R6
($-2.5$~percentage points), the overall exclusion rates differ by only $2.8$~percentage
points ($58.2\%$ synthetic vs.\ $55.4\%$ empirical). This partial
cancellation occurs because the more aggressive synthetic R5 exclusion is
offset by the weaker synthetic R6 exclusion. The near-agreement at the
overall level suggests that, for catalog-scale demographics, the specific
proxy choice has a modest effect on the aggregate exclusion fraction.
However, the \textit{identity} of excluded stars differs substantially: the
synthetic approach excludes different individual stars than the empirical
approach, which matters for target-by-target scheduling decisions.
We therefore recommend using empirical flags wherever available, supplemented
by RUWE as an additional conservative filter for multiplicity.

\subsection{Cross-Match with Breakthrough Listen}
\label{sec:crossmatch}

\begin{figure}[!h]
    \includegraphics[width=\columnwidth]{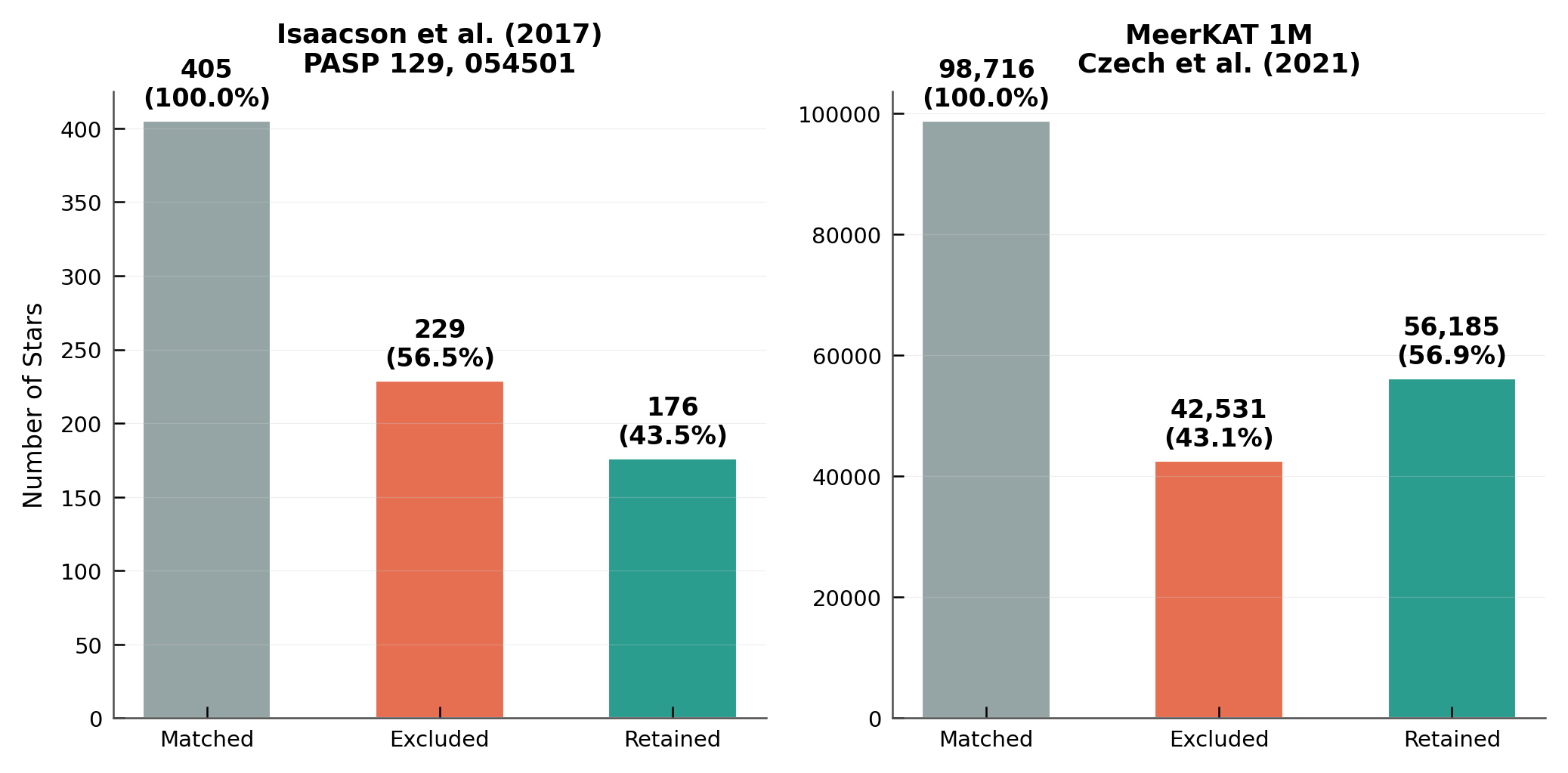}
    \caption{Cross-match results with the Breakthrough Listen primary target list}
    \citep{Isaacson2017} and the MeerKAT 1M sample \citep{Czech2021}.
    The higher exclusion rate for BL primaries reflects the complementary
    selection philosophies: BL prioritizes detectability (nearby, bright),
while the Torlakcik Catalog prioritizes habitability (old, metal-rich, stable).
    Isaacson coordinates were propagated to J2016.0 using proper motions prior
    to cross-matching with a $5''$ radius; MeerKAT was matched via \texttt{source\_id}.
    \label{fig:bl_crossmatch}
\end{figure}

To contextualize this catalog relative to operational SETI programs, we
cross-matched against two Breakthrough Listen target samples: the primary
target list of \citet{Isaacson2017} ($N = 1,709$ stars) and the MeerKAT
1M sample of \citet{Czech2021} ($N \approx 1.2$ million stars).
For the Isaacson catalog, coordinates were propagated from the catalog epoch
(J2000, with per-star epoch information from the \texttt{Ep} column) to the
Gaia DR3 reference epoch (J2016.0) using the proper motion values
(\texttt{pmRA}, \texttt{pmDE}; confirmed as arcsec/yr from the FITS TUNIT
header) prior to cross-matching, accounting for the $\sim$16-year epoch
difference that can shift nearby high-proper-motion stars by tens of
arcseconds. A $5''$ matching radius was then applied. For the MeerKAT
sample, which includes \texttt{source\_id}, we performed a direct
\texttt{source\_id} cross-match, eliminating any positional uncertainty.
Results are summarized in Table~\ref{tab:crossmatch} and
Figure~\ref{fig:bl_crossmatch}, and discussed in
Section~\ref{sec:bl_discussion}.

\begin{table}[!h]
\centering
\caption{Cross-match results with Breakthrough Listen surveys.}
\label{tab:crossmatch}
\begin{tabular}{lrrrr}
\toprule
Sample & Total & Matched & Excluded & Retained \\
\midrule
BL Primary \citep{Isaacson2017} & 1,709 & 405 & 229 (56.5\%) & 176 (43.5\%) \\
MeerKAT 1M \citep{Czech2021} & $\sim$1.2M & 98,716 & 42,531 (43.1\%) & 56,185 (56.9\%) \\
\bottomrule
\end{tabular}
\end{table}

\begin{figure}[!h]
    \centering
    \includegraphics[width=\columnwidth]{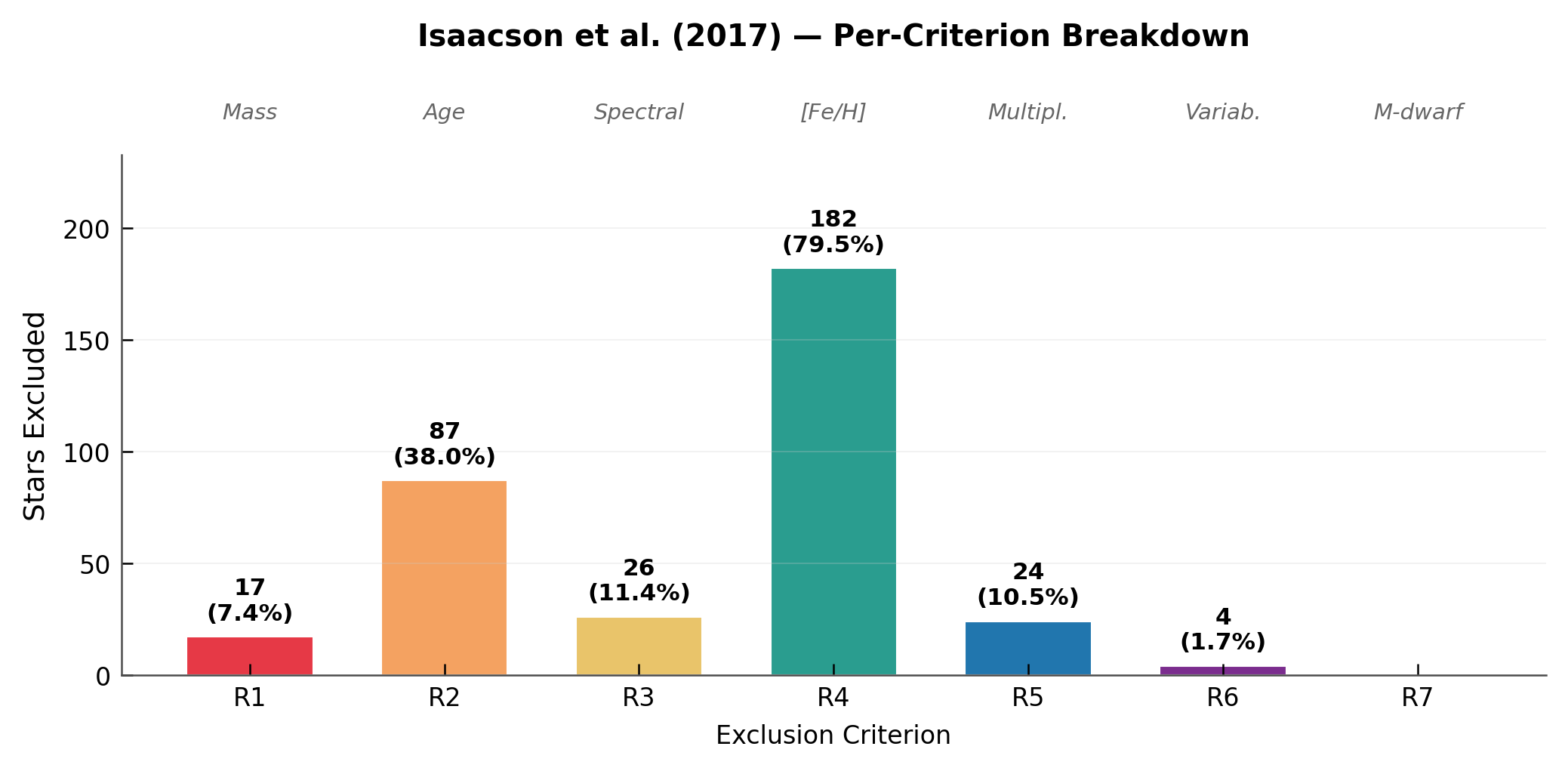}
    \caption{Per-criterion exclusion breakdown for the Isaacson et al.\ (2017)
    BL primary sample. Each bar shows the number of excluded targets attributable
    to each criterion, with percentages relative to the total excluded count.
    Low metallicity (R4) is the dominant exclusion driver, reflecting the metallicity distribution of the matched subsample}
    \label{fig:bl_criteria_isaacson}
\end{figure}

\begin{figure}[!h]

    \includegraphics[width=\columnwidth]{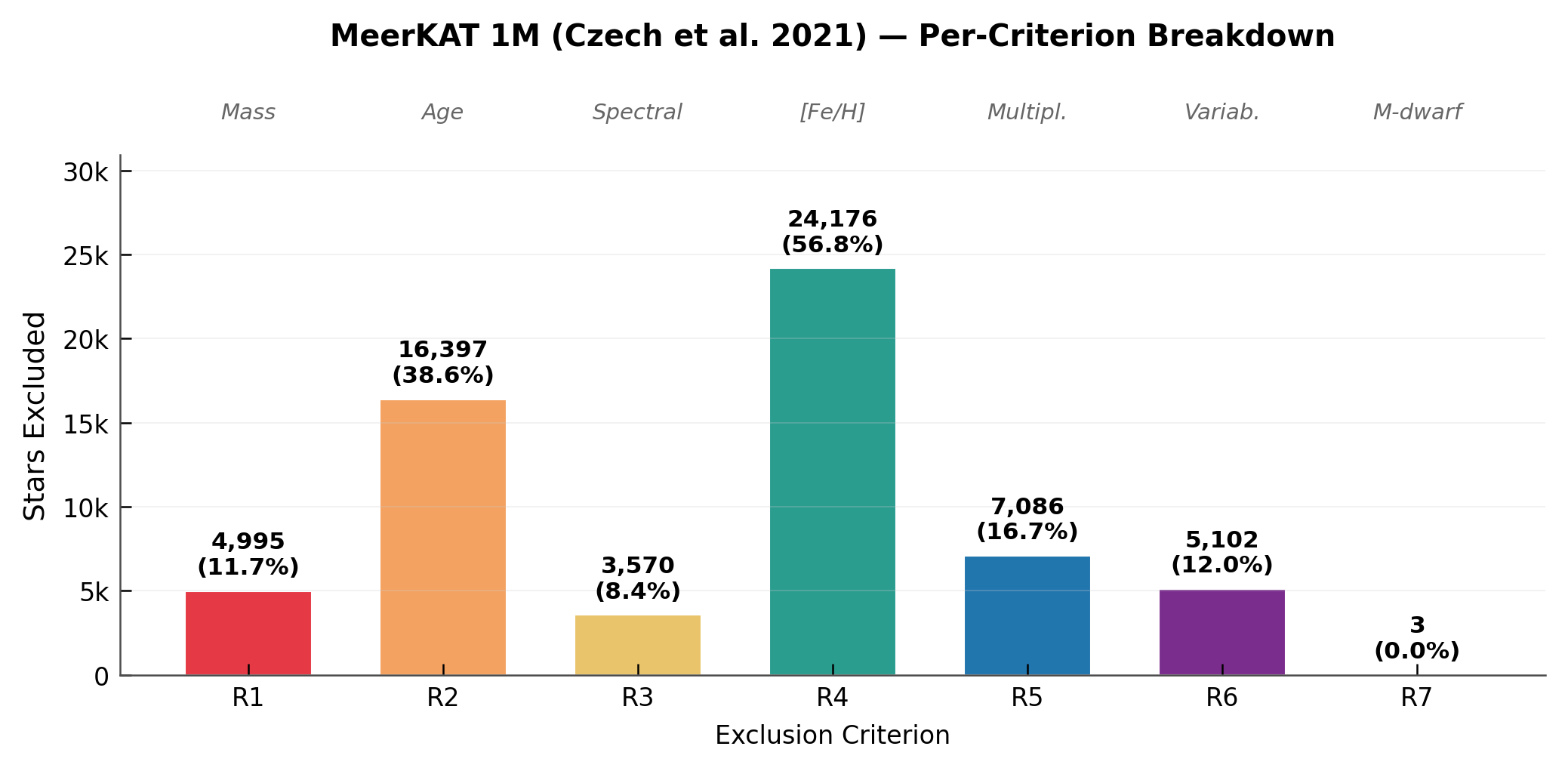}
    \caption{Same as Figure~\ref{fig:bl_criteria_isaacson} but for the
    MeerKAT 1M sample \citep{Czech2021}. The much larger sample size
    yields a different criterion distribution, with R5 (multiplicity) and
    R6 (photometric variability) contributing a larger fraction of
    exclusions than in the targeted Isaacson sample.}
    \label{fig:bl_criteria_meerkat}
\end{figure}

\subsection{Sky Distribution}

\begin{figure}[!h]
 
      \includegraphics[width=\columnwidth]{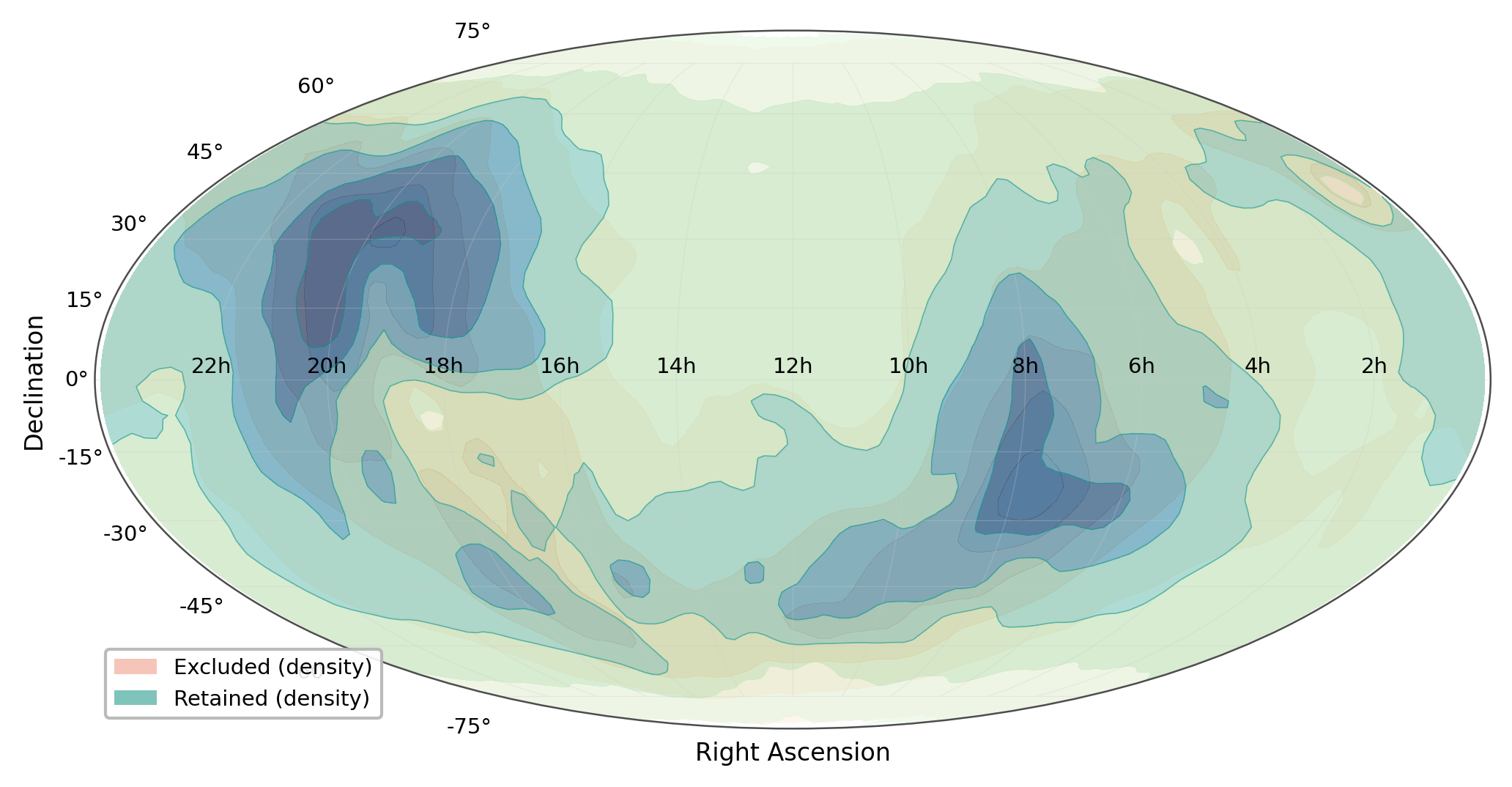}
    \caption{All-sky density distribution of the Gaia DR3 sample
    ($N = 1{,}742{,}306$). Orange contours: excluded star density.
    Teal contours: retained candidate density. Contour levels are
    proportional to surface density. No systematic spatial bias in
    the exclusion pattern is apparent beyond the Galactic plane
    stellar population gradient.}
    \label{fig:skymap}
\end{figure}

Figure~\ref{fig:skymap} shows the all-sky distribution of retained and excluded
stars. No systematic spatial bias in the exclusion pattern is apparent beyond the
stellar population gradient near the Galactic plane.

\section{Sensitivity Analysis}
\label{sec:sensitivity}

\begin{figure}[h]
    \centering
    \includegraphics[width=\columnwidth]{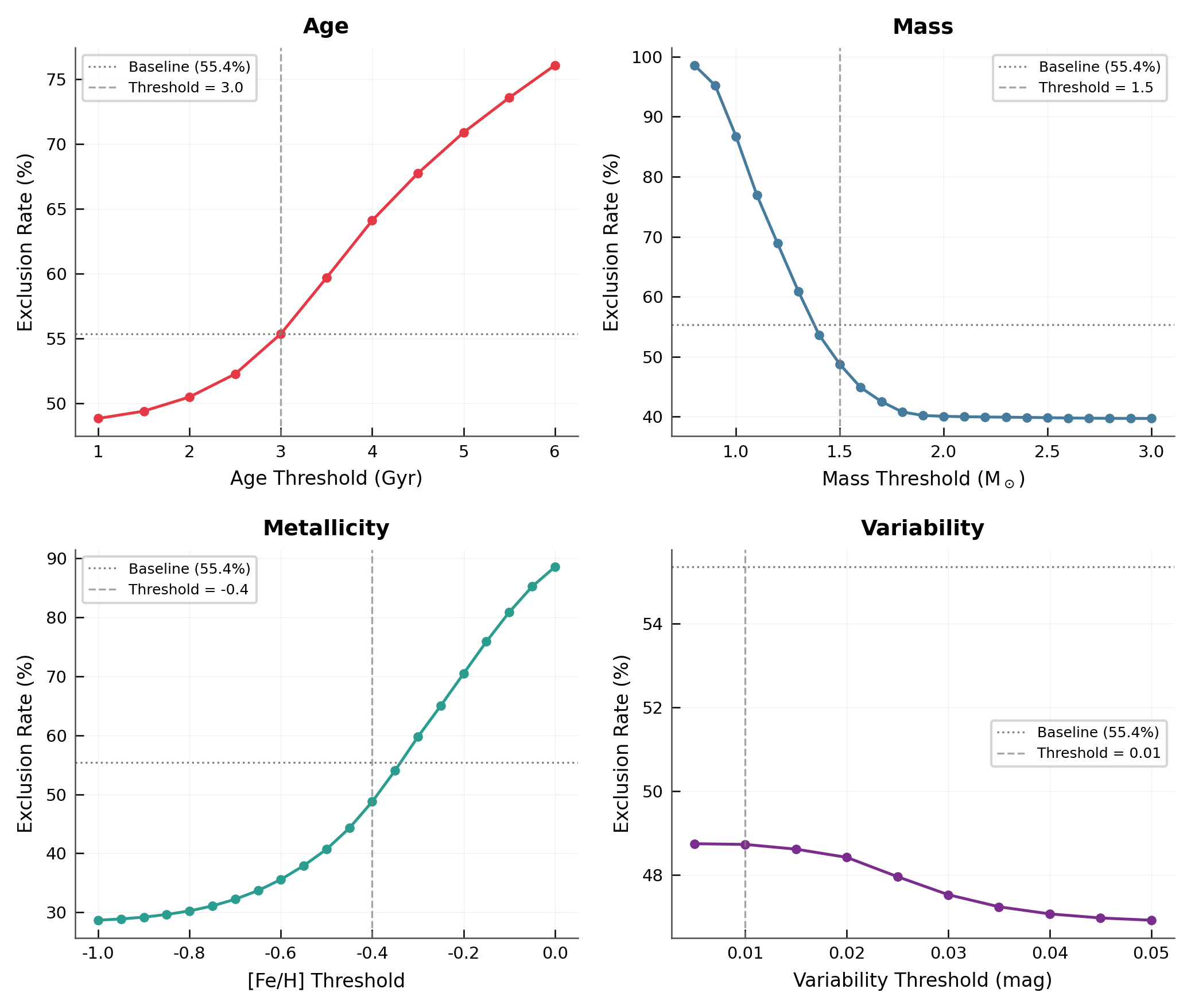}
    \caption{Sensitivity of the overall exclusion rate to threshold variations
    for age, mass, metallicity, and variability. Dashed vertical lines mark
    baseline thresholds; the horizontal dashed line indicates the baseline
    exclusion rate (55.4\%). The age threshold dominates the sensitivity,
    while the variability threshold has negligible effect.}
    \label{fig:sensitivity}
\end{figure}

We performed a one-at-a-time sensitivity analysis, varying each continuous
threshold while holding all others fixed at their baseline values
(Figure~\ref{fig:sensitivity}).

\textbf{Age threshold} (baseline: 3~Gyr) is the dominant driver. Varying it
from 1 to 6~Gyr moves the overall exclusion rate from $48.8\%$ to $76.1\%$.
The inflection in the curve near 2.5~Gyr (where the exclusion rate is $52.3\%$)
supports the physically motivated choice of 3~Gyr following \citet{Turnbull2003}
rather than an arbitrary selection: the steepest gradient in exclusion rate
occurs across the 2.5--3.5~Gyr range, where a 1~Gyr change in threshold shifts
the exclusion rate by $\sim$7~percentage points.

\textbf{Mass threshold} (baseline: $1.5\,M_\odot$) shows a clear knee at the
baseline value. Below $1.5\,M_\odot$, the exclusion rate increases steeply
(from $53.9\%$ at $1.9\,M_\odot$ to $98.6\%$ at $0.8\,M_\odot$), while above
$1.5\,M_\odot$ the curve flattens: the exclusion rate at $1.9\,M_\odot$ is
essentially identical to that at $3.0\,M_\odot$ ($53.9\%$). This tells us that
few stars in the sample have masses between 1.5 and $3.0\,M_\odot$ that are not
already excluded by other criteria, and that the baseline threshold sits
right at the transition.

\textbf{Metallicity threshold} (baseline: $[\mathrm{Fe/H}] = -0.4$) shows
monotonically increasing exclusion as the threshold moves toward solar
metallicity, reflecting the roughly Gaussian metallicity distribution of
the sample. The exclusion rate increases from $39.1\%$ at
$[\mathrm{Fe/H}] = -1.0$ to $88.9\%$ at solar metallicity, with the steepest
gradient near the baseline value.

\textbf{Variability threshold} (baseline: 0.01~mag) has the smallest effect,
with total exclusion varying by only $\sim$1.7~percentage points across the
tested range (0.005--0.05~mag). The exclusion rate decreases from $55.4\%$ at
0.005~mag to $53.7\%$ at 0.05~mag. This insensitivity confirms that the R6
criterion, while physically motivated, is a minor contributor to the overall
exclusion budget regardless of the exact threshold chosen.

\section{Discussion}
\label{sec:discussion}

\subsection{Intended Users}

This catalog is designed for three primary user communities. First, \textit{small-telescope
SETI programs} and citizen science initiatives with limited resources for complex target
selection pipelines can use the pre-filtered catalog directly as an input target list.
Second, \textit{wide-field survey programs} such as MeerKAT or the VLA benefit from
the sky density analysis: with 18.86 retained stars per square degree and a mean
separation of $\sim$14~arcmin, telescopes with beams wider than $\sim$15~arcmin have
 $> 50\%$ probability of finding at least one retained candidate in the primary beam
at any pointing (Table~\ref{tab:fov}). 
The by-catch principle \citep{Isaacson2017}, that is, pointing at a formally lower-priority target still observes nearby higher-priority stars within the beam, operates efficiently at this sky density. Third, \textit{catalog pipeline
developers} can apply the avoidance function to any future stellar catalog (Gaia DR4,
PLATO input catalogs, etc.) without modification, using the publicly available
\textit{stellar-avoidance} community software that provides a CLI interface,
configurable criteria, and full provenance tracking for audit-ready reproducibility.
\begin{table}[!h]
\centering
\caption{Field-of-view coverage statistics for major SETI-relevant facilities
at 1.4~GHz. Sky density computed from 777,835 retained stars over the full sky.}
\label{tab:fov}
\begin{tabular}{lrrr}
\toprule
Telescope & Beam FWHM & $\lambda_\mathrm{expected}$ & $P(\geq 1)$ \\
          & (arcmin)  & (stars/beam)                 &             \\
\midrule
GBT       &  9.0 & 0.33 & 28.3\% \\
Parkes    & 14.0 & 0.81 & 55.3\% \\
VLA       & 30.0 & 3.70 & 97.5\% \\
MeerKAT   & 58.0 & 13.84 & $\sim$100\% \\
\bottomrule
\end{tabular}
\end{table}

\subsection{Relationship to Breakthrough Listen and Existing Surveys} \label {sec:bl_discussion}

To contextualize our habitability filter relative to operational SETI programs, we
cross-matched the Torlakcik Catalog against two Breakthrough Listen target
samples.

The \textit{BL primary sample} \citep{Isaacson2017} comprises 1,709 nearby stars
selected for targeted observations with the Green Bank Telescope and Parkes.
This sample prioritizes proximity (typically within 50~pc) and observability.
Because the Isaacson catalog reports coordinates at the Hipparcos epoch
($\sim$J2000) with proper motions (\texttt{pmRA}, \texttt{pmDE} in
arcsec/yr, as confirmed from the FITS TUNIT header), we propagated all
coordinates to the Gaia DR3 reference epoch (J2016.0) prior to
cross-matching. This correction is essential for nearby high-proper-motion
stars: the median positional shift is $4.7''$, with 795 stars shifting by
more than $5''$ and the maximum reaching $165.8''$ (Barnard's Star).
After PM correction, cross-matching with a $5''$ radius recovered 405 of the
1,709 BL primary targets in our Gaia DR3 sample (23.7\%), an 88\%
increase over the 215 matches obtained without PM correction. The low match rate reflects our Gaia DR3 query, which requires non-null GSP-Phot and FLAME parameters. Many bright nearby BL stars saturate in Gaia ($G \lesssim 6$~mag) and lack these derived parameters. These stars are absent from our sample entirely. They were not rejected by the avoidance criteria; they simply lack the input data needed for evaluation. Of the 405 matched
stars, 229 (56.5\%) are flagged for exclusion by our model, with low
metallicity (R4) accounting for 182 of the exclusions. This result reflects
a deliberate difference in survey philosophy: BL selects stars where
technosignatures are most \textit{detectable} (nearby, bright), while our model
selects stars where they are most \textit{plausible} (old, metal-rich, stable).
The two approaches are complementary: applying our habitability filter to the
BL primary list would retain 176 targets that satisfy both selection criteria.

The \textit{MeerKAT sample} \citep{Czech2021} is a substantially larger target
list developed for commensal SETI observations during MeerKAT large survey
programs. The complete sample contains approximately 1.2 million stars drawn
from Gaia DR2 with data quality filters and expected pointing positions.
Cross-matching yields 98,716 stars present in both catalogs, of which 42,531
(43.1\%) are excluded and 56,185 (56.9\%) retained by our model. The lower
exclusion rate compared to the BL primary sample reflects the MeerKAT sample's
broader selection function, which includes more distant stars with higher
median metallicity. The non-detection rate (fraction of targets not found in our catalog) is higher for MeerKAT than for BL primaries (91.8\% vs.\ 76.3\%), again because most MeerKAT targets lack the astrophysical parameters our query requires. With a 58-arcmin beam, MeerKAT observes an expected
 $\sim$14 retained stars per pointing, meaning most pointings include retained candidates even when observing other targets.

The low match rate between BL and our sample does not limit the tool's practical utility. The publicly released stellar-avoidance software allows users to enable or disable individual criteria and adjust all thresholds. A survey that needs full coverage of a specific input catalog can simply disable criteria that depend on parameters missing from that catalog. Most of the unmatched BL stars are bright sources that saturate in Gaia DR3 and will receive GSP-Phot and FLAME parameters in Gaia DR4, which will substantially close this coverage gap.

These cross-match results show that the Torlakcik Catalog can serve as
an upstream habitability filter: for narrow-field targeted surveys (GBT, Parkes),
it provides a scientifically motivated subset of existing lists; for wide-field
commensal surveys (MeerKAT, VLA), it confirms sufficient target density for
practical survey design.

\subsection{Implications of Fractional Flux Error as a Variability Proxy}

The result that $\sigma_G/G$ is an insensitive proxy for photometric variability
(Section~\ref{sec:proxy}) deserves further discussion, as it runs counter to the
intuitive expectation that variable stars should exhibit larger measurement
uncertainties. The resolution lies in recognizing that Gaia's mean photometry
is not a simple average of flux measurements but rather the output of a
forward model that accounts for the satellite's scanning law, CCD gate
timing, and background subtraction \citep{Riello2021}. For stars with many
observations (the typical Gaia source has $\sim$40--250 transits in DR3), the
mean flux converges to a precise value even when individual measurements vary
substantially. The error bar on the mean therefore reflects the precision of
the \textit{average}, not the \textit{scatter} of the measurements. The
correct quantity for assessing variability amplitude is the range or standard
deviation of individual photometric measurements, which is precisely what
\texttt{range\_mag\_g\_fov} and the Gaia variability classification pipeline
provide. This distinction between precision-of-the-mean and
amplitude-of-variation is well-known in statistical practice but has, to our
knowledge, not been explicitly demonstrated in the context of SETI target
selection. We emphasize this finding because fractional photometric error is
an easily computed and intuitively appealing quantity; our results show that
it should not be used as a variability indicator without explicit validation.

\subsection{Rule-Based vs.\ Probabilistic Approaches}

We chose a rule-based design over a probabilistic one. A logistic regression or
Bayesian ranking model would likely be slightly more accurate in recovering
HabCat-like targets, but it would make it harder to understand why any individual
star was rejected. For scheduling pipelines where rejection decisions need to be
traceable, the interpretability and low computational cost of rule-based approaches
are practical advantages.

A natural probabilistic extension would assign continuous compatibility scores
rather than binary decisions:
\begin{equation}
s_i = \begin{cases}
1 - M/1.5\,M_\odot & \text{(mass)} \\
\tau_\mathrm{upper}/3\,\mathrm{Gyr} & \text{(age)} \\
([\mathrm{Fe/H}] + 0.4)/0.4 & \text{(metallicity)}
\end{cases}
\end{equation}
A composite score $S = \min_i(s_i)$ would rank targets continuously; the
threshold $S > 0$ recovers the binary output of this work while providing finer
prioritization within the retained set. We defer this extension to future work.

\subsection{Relationship to HabCat}

The selection criteria used here closely follow those of \citet{Turnbull2003}.
HabCat is no longer publicly available in machine-readable form, which prevented
a direct cross-match; however, the astrophysical justifications for each criterion
come from the same literature. The main methodological differences are: (1) our
model is catalog-agnostic; (2) it treats M-dwarf chromospheric activity as a
separate criterion; (3) it applies the age cut to the upper bound rather than
the point estimate; and (4) it releases a complete machine-readable exclusion
catalog with reason codes.

\subsection{M-Dwarf Treatment}

The $91\%$ exclusion rate for M dwarfs is driven mainly by the age criterion (R2)
and metallicity (R4), not by chromospheric activity (R7 flags only 84 stars).
In a volume-limited program the fraction of old, quiet M dwarfs would be
considerably higher. The model correctly retains chromospherically quiet M dwarfs as candidates,
consistent with \citet{Shields2016}. \citet{Basri2026} notes that red dwarf
magnetic activity eventually decreases with age, suggesting that old, inactive
M dwarfs may be more hospitable than their activity levels during youth would
imply.
However, all M dwarfs experience elevated magnetic activity during their first few Gyr, producing flares, UV, and X-ray emission. Because the habitable zone lies at only $\sim$0.1~au, any terrestrial planet receives concentrated doses of this radiation. Recent atmosphere searches with JWST for HZ exoplanets around M dwarfs have returned null results, reinforcing the concern that M-dwarf HZ planets may lack atmospheres entirely. Users who wish to exclude all M dwarfs can enable the spectral type criterion (R3) for M-type stars via the configurable community tool.

\subsection{False Positive Mitigation in Radio SETI: Background Confusion}

The primary observational modality for the targets retained in this catalog is radio SETI. A persistent practical challenge in single-dish and wide-field radio surveys is signal confusion and increased system noise ($T_\mathrm{sys}$) resulting from background extragalactic radio sources, such as Active Galactic Nuclei (AGNs). If a retained SETI target lies along the same line of sight as a strong background continuum emitter, the target may yield false-positive technosignature triggers or suffer from severely degraded observational sensitivity.

While the current Torlakcik Catalog strictly filters based on stellar astrophysical parameters, a highly practical future extension involves cross-matching the retained candidate list with major radio continuum catalogs (e.g., NVSS, VLASS). Assigning a ``Radio Confusion Flag'' to stars that fall within the beam solid angle of a known $>10$~mJy background source would provide radio observatories with an immediate, pre-calculated metric of observational clarity. This would ensure that valuable integration time at facilities like the GBT or MeerKAT is prioritized for targets with not only the right stellar physics but also clean radio backgrounds.

\subsection{Limitations}
\label{sec:limitations}
Several limitations should be noted. First, the parallax filter
($\varpi/\sigma_\varpi > 5$) introduces a distance-dependent completeness
boundary; the sample is neither volume-limited nor magnitude-limited. In
practice, though, the filter removed only $11{,}829$ stars ($0.67\%$) from
the initial query, indicating that this boundary has negligible effect on
the aggregate exclusion statistics reported here, while remaining a
consideration for completeness-sensitive analyses. Second, the \texttt{non\_single\_star} flag may
miss unresolved close binaries that lack NSS solutions. Third, the M-dwarf activity
criterion uses Gaia variability flags designed for general variability classification
rather than flare detection; dedicated flare catalogs from TESS or Kepler would improve
coverage. Fourth, photometric variability and M-dwarf activity flags (R6, R7) are
available for only 66,929 of the $1{,}742{,}306$ stars; the remaining sources were
conservatively treated as non-variable, likely underestimating the true R6/R7
exclusion contributions. Fifth, Gaia photometric ages carry $\sim$$40\%$ fractional
uncertainties for individual stars, with typical uncertainties of
 $\sim$2.5~Gyr \citep{GaiaCollaboration2023}. The FLAME algorithm
also shows a known bias toward younger ages for low-mass stars; the
upper-bound criterion mitigates but does not eliminate this limitation.

\section{Conclusions}
\label{sec:conclusions}

We have presented the \textit{Torlakcik Catalog}, a parametric avoidance
model for SETI target selection applied to a Gaia DR3 sample of
$N = 1{,}742{,}306$ stars with reliable parallax. The main results are:

\begin{enumerate}
  \item The model excludes $\sim$55$\%$ of candidates and retains $\sim$778,000 ($\sim$44$\%$)
    high-priority targets. Age and metallicity are co-dominant exclusion drivers,
    each affecting $\sim$29$\%$ of stars.

  \item Applying the age threshold to Gaia DR3 age upper bounds rather than point
    estimates retains 355,086 additional candidates ($+41\%$) relative to a hard cut,
    accounting for the large photometric age uncertainties.

  \item A systematic comparison of empirical Gaia DR3 classification flags against
    synthetic proxies reveals that: (a) RUWE $>$ 1.4 flags 2.7 times more stars for
    multiplicity than the Gaia non-single-star flag ($18.4\%$ vs.\ $6.9\%$), providing
    a more conservative filter for SETI purposes; (b) the fractional G-band flux error
    $\sigma_G/G$ is an insensitive proxy for photometric variability, capturing only
    $36\%$ of empirically identified variables ($1.4\%$ vs.\ $3.9\%$) due to Gaia's
    extraordinary mean-flux precision (median $\sigma_G/G = 0.020\%$); and (c) the
    overall exclusion rates agree to within 2.8~percentage points ($58.2\%$ synthetic
    vs.\ $55.4\%$ empirical) because R5 overestimation and R6 underestimation
    partially cancel. We recommend using empirical flags where available,
    supplemented by RUWE as an additional conservative filter for multiplicity.

  \item Cross-matching with the Breakthrough Listen primary list reveals that $56.5\%$
    of matched BL targets would be excluded by our habitability criteria, primarily due
    to low metallicity. This reflects complementary selection philosophies rather than
    an inconsistency. We note that only 405 of 1,709 BL targets were matched,
    limiting the generality of this comparison.

  \item The median distance of retained candidates is $382$~pc, within the sensitivity
    volume of current radio SETI facilities. Sky density ($\sim$$19$~stars/sq~deg)
    is sufficient that any pointing with MeerKAT or VLA contains multiple
    retained candidates in the primary beam.

  \item The framework is catalog-agnostic, lightweight, and produces machine-readable
    reason codes suitable for integration into observational scheduling pipelines.
    Both the paper-specific analysis pipeline and a generalized, reproducible,
    audit-ready community software are publicly available, enabling any SETI program
    to apply the avoidance model to arbitrary stellar catalogs with full
    provenance tracking and configurable criteria.
\end{enumerate}

\section{Data Availability}
\label{sec:data_availability}

Two complementary software products are publicly released:

\begin{itemize}
\item The \textit{paper-specific analysis pipeline}, including all data-processing
  modules, figure generation scripts, and cross-match routines used to produce
  the results reported in this paper, is publicly available at
  \url{https://github.com/torlakciksahin/gaia-seti-avoidance}.

\item The \textit{stellar-avoidance} community software, a generalized,
  catalog-agnostic, audit-ready Python package with configurable criteria,
  CLI interface, and full provenance tracking, is publicly available at
  \url{https://github.com/torlakciksahin/stellar-avoidance}.
\end{itemize}

The \textit{Torlakcik Catalog}, a machine-readable Gaia DR3
exclusion catalog containing $N = 1{,}742{,}306$ stars, is
archived on Zenodo at
\url{https://doi.org/10.5281/zenodo.19956677}.

The underlying Gaia DR3 data were accessed via the ESA TAP
service; the complete TAP query used to construct the catalog
is reproduced in Appendix~\ref{app:query}.
%bestpart:)

\section*{Acknowledgments}
This work was carried out independently by the author as a high school student, without any external funding or institutional support.
The author expresses gratitude to Jason T. Wright (Penn State) for providing detailed and constructive feedback that substantially improved this manuscript.
The author also thanks Howard Isaacson (UC Berkeley) for his detailed and insightful feedback.
This work made use of data from the European Space Agency mission Gaia, processed by the Gaia Data Processing and Analysis Consortium (DPAC). The author used Gemini 3 solely for language polishing and improving the readability of the manuscript. All scientific content, analysis, results, and conclusions are the author's own work. The author takes full responsibility for the final text.

\appendix

\section{Gaia DR3 TAP Query}
\label{app:query}

\begin{verbatim}
SELECT
  gs.source_id, gs.ra, gs.dec,
  gs.parallax, gs.parallax_over_error,
  gs.phot_g_mean_mag, gs.bp_rp,
  ap.teff_gspphot, ap.mh_gspphot,
  ap.logg_gspphot,
  fl.mass_flame_spec,
  fl.age_flame_spec,
  fl.age_flame_spec_upper,
  gs.non_single_star,
  gs.phot_variable_flag,
  vs.range_mag_g_fov,
  vs.in_vari_rotation_modulation,
  vs.in_vari_short_timescale,
  gs.ruwe,
  gs.phot_g_mean_flux,
  gs.phot_g_mean_flux_error
FROM gaiadr3.gaia_source AS gs
JOIN gaiadr3.astrophysical_parameters AS ap
  ON gs.source_id = ap.source_id
JOIN gaiadr3.astrophysical_parameters_supp AS fl
  ON gs.source_id = fl.source_id
LEFT JOIN gaiadr3.vari_summary AS vs
  ON gs.source_id = vs.source_id
WHERE ap.teff_gspphot IS NOT NULL
AND ap.mh_gspphot IS NOT NULL
AND fl.mass_flame_spec IS NOT NULL
AND fl.age_flame_spec IS NOT NULL
AND gs.parallax_over_error > 5
\end{verbatim}

\label{app:codes}

\begin{table}[H]
\centering
\caption{Machine-readable reason codes assigned by the parametric avoidance function.}
\label{tab:codes}
\begin{tabular}{lll}
\toprule
Code & Criterion & Threshold \\
\midrule
R1 & Stellar mass       & $> 1.5\,M_\odot$ \\
R2 & Stellar age (upper bound) & $< 3$~Gyr \\
R3 & Spectral type      & O, B, A, F0--F4 \\
R4 & Metallicity        & $[\mathrm{Fe/H}] < -0.4$ \\
R5 & Multiplicity       & \texttt{non\_single\_star} $\geq 1$ \\
R6 & Phot.\ variability & $>0.01$~mag or VARIABLE flag \\
R7 & M-dwarf activity   & rotation/short-timescale flag \\
RETAIN & ---            & No criterion satisfied \\
\bottomrule
\end{tabular}
\end{table}

\bibliography{Bibliography}
\bibliographystyle{aasjournal}
 
\end{document}